\documentclass{aa}  
\bibpunct{(}{)}{;}{a}{}{,} 

\usepackage{graphicx}
\usepackage[varg]{txfonts}
\begin{document}

\title{Precise mass and radius measurements for the components of the bright
solar-type eclipsing binary star V1094 Tau\thanks{The light curves shown in
Fig.~1 are available in electronic form at
the CDS via anonymous ftp to cdsarc.u-strasbg.fr (130.79.128.5) or via
http://cdsweb.u-strasbg.fr/cgi-bin/qcat?J/A+A/.}}

\titlerunning{V1094 Tau}
\author{P.~F.~L.~Maxted\inst{1} 
\and
R.~J.~Hutcheon\inst{1} 
\and
G.~Torres\inst{2}
\and
C.~H.~S.~Lacy\inst{3}
\and
J. Southworth\inst{1}
\and
B. Smalley\inst{1} 
\and
K. Pavlovski\inst{4}
\and 
\\
L. A. Marschall\inst{5}
\and 
J.~V.~Clausen\inst{6}}           
          
\institute{Astrophysics Group,  Keele University, Keele, Staffordshire,
ST5~5BG, UK\\
\email{p.maxted@keele.ac.uk, richard.hutcheon@btinternet.com}
\and
Harvard-Smithsonian Center for Astrophysics, 60 Garden Street, Cambridge,
MA 02138, USA\\
\email{gtorres@cfa.harvard.edu,  rstefanik@cfa.harvard.edu}
\and
Physics Department, University of Arkansas, Fayetteville, AR 72701, USA \\
\email{clacy@uark.edu }
\and
Department of Physics, University of Zagreb, Bijeni\v{c}ka cesta 32, 10000
Zagreb, Croatia \\
\email{pavlovski@phy.hr }
\and
Department of Physics, Gettysburg College, 300 North Washington Street,
Gettysburg, PA 17325, USA\\
\email{marschal@gettysburg.edu}
\and
Niels Bohr Institute, Copenhagen University, Juliane Maries Vej 30,
DK-2100 Copenhagen \/{O}, Denmark\thanks{Deceased 2011 June 5}
}

\date{Dates to be inserted}

 
  \abstract
{V1094 Tau is bright eclipsing binary star with an orbital period close to 9
days that contains two stars similar to the Sun. }
{Our aim is to test models of Sun-like stars using precise and accurate
mass and radius measurements for both stars in V1094~Tau.}
{We present new spectroscopy of V1094 Tau, which we use to estimate the
effective temperatures of both stars and to refine their spectroscopic orbits.
We also present new, high-quality photometry covering both eclipses of V1094
Tau in the Str\"{o}mgren {\it uvby} system and in the Johnson V-band.}
{The masses,  radii, and effective temperatures of the stars in V1094~Tau are
found to be $M_A = 1.0965 \pm 0.0040\,M_{\sun} $, $R_A = 1.4109 \pm
0.0058\,R_{\sun}$, T$_{\rm eff, A} = 5850 \pm 100$\,K, $M_B = 1.0120 \pm
0.0028\,M_{\sun} $, $R_B = 1.1063 \pm 0.0066\,R_{\sun}$, and T$_{\rm eff, B} =
5700 \pm 100$\,K. An analysis of the times of mid-eclipse and the radial
velocity data reveals apsidal motion with a period of $14500\pm 3700$
years.}
{The observed masses, radii, and effective temperatures are consistent with
stellar models for an age $\approx 6$\,Gyr if the stars are assumed to have a
metallicity similar to the Sun. This  estimate is in reasonable agreement with
our estimate of the metallicity derived using Str\"{o}mgren photometry and
treating the binary as a single star ([Fe/H]$=-0.09\pm0.11$). The rotation
velocities of the stars suggest that V1094~Tau is close to the limit at which
tidal interactions between the stars force them to rotate pseudo-synchronously
with the orbital motion.}

\maketitle
%

\section{Introduction}

\object{V1094 Tau} (HD 284195) is a ninth-magnitude star that was discovered to
be an eclipsing binary star by  \citet{1994IBVS.4119....1K} using photographic
photometry. The period quoted  in that paper is incorrect and was later found
to be the time between the secondary and primary minima. The correct period
($P\approx8.988$\,d) was first given by \citet{1998IBVS.4544....1K}, who also
established that the orbit is eccentric, with the secondary minimum falling
near phase 0.65. Observations of the times of mid-eclipse have continued since
its discovery and have been analysed by \citet{2010A+A...509A..18W}, who
claim to have detected apsidal motion with a period of 13600 years and
variations in these times of mid-eclipse  with a period of about eight years and
an amplitude of 200\,s that they suggest may be due to a third body in the
system. 

 The spectroscopic orbits of both stars have been measured to good accuracy by
\citet{2003Obs...123..203G}. No good light curves of the star were available at
that time, but they were able to establish that the rotation of the stars is
likely to be pseudo-synchronised with the orbital motion, that the masses are
close to 1.1\,$M_{\sun}$ and 1.0\,$M_{\sun}$, and that the spectral types of
the stars are approximately G0 and G3.  

 The techniques for  measuring the masses and radii of the stars in eclipsing
binaries, such as V1094~Tau to good accuracy ($\approx 1$\%) independently of
any stellar structure models, are now well established. To obtain masses and
radii to this precision requires light curves of good quality that cover both
eclipses and spectroscopic orbits with good phase coverage based on
high-resolution spectroscopy. With data of this quality, it is possible to
critically test stellar structure models, particularly if these mass and
radius estimates can be complemented with reliable estimates for the effective
temperatures of the stars and their metallicity \citep{2010A+ARv..18...67T}.
Compilations of such data have also been used to establish empirical
relationships between mass, radius, effective temperature, etc.  These
empirical relationships can be used, for example, to estimate the mass of a
star based on its observed density, effective temperature, and metallicity.
These quantities are directly measurable for stars in transiting exoplanet
systems \citep{2010A+A...516A..33E}.

 In this work we present the first high-quality light curves and the first
high-resolution spectroscopy for V1094~Tau. We use these data to measure the
masses and radii of the stars in this binary to better than 1\%. We also make
useful estimates of the effective temperatures of the stars and their
metallicity based on Str\"omgren photometry.

\section{Observations}
 
\subsection{Spectroscopy}
 V1094~Tau was observed with a Cassegrain-mounted echelle spectrograph
attached to the 1.5-m Wyeth reflector at the Oak Ridge Observatory 
(Harvard, Massachusetts, USA). A total of 59 useful exposures were obtained
between 1995 December and 1997 April at a resolving power of R $\approx$
35,000. A single echelle order was recorded with an intensified
photon-counting Reticon detector, giving 45\AA\ of coverage centred at
5187\AA, and including the lines of the Mg\,{\sc i}\,b triplet. The
signal-to-noise ratios range from 13 to 24 per resolution element of
8.5\,km\,s$^{-1}$. The wavelength calibration was established by means of
exposures of a Th-Ar lamp before and after each science exposure, and all
reductions were carried out with standard procedures as implemented in a
dedicated pipeline \citep[see][]{1992ASPC...32..110L}.

 Spectroscopic observations were also obtained in 2002 October using the 2.5-m
Isaac Newton Telescope (INT) on La Palma. The 500\,mm camera of the
Intermediate Dispersion Spectrograph (IDS) was equipped with a holographic
2400\,lines\,mm$^{-1}$ grating. An EEV 4k\,$\times$\,2k CCD was used and
exposure times were 300\,s. From measurements of the full width at half
maximum (FWHM) of arc lines taken for wavelength calibration, we estimated
that the resolution is 0.2\,\AA. A total of 64 spectra were taken covering the
interval 4230--4500\,\AA, with estimated signal-to-noise ratios of
approximately 50 per pixel. 

The reduction of all spectra was undertaken using optimal extraction
\citep{1986PASP...98..609H} as implemented in the software tools {\sc pamela}
and {\sc molly}\footnote{{\sc pamela} and {\sc molly} were written by Prof.
Tom Marsh and are available at \url{www.warwick.ac.uk/go/trmarsh}}
\citep{1989PASP..101.1032M}.

\subsection{Photometry}

 The differential $uvby$ light curves of V1094~Tau were observed at the
Str{\"o}mgren Automatic Telescope (SAT) at ESO, La Silla and its six-channel
$uvby\beta$ photometer on 72 nights between October 2000 and January 2008.
They contain 670 points per band. HD\,26736  and HD\,26874  were used as
comparison stars throughout. In addition HD\,27989AB was used as comparison
star until JD2451889, but was found to be variable and was therefore replaced
by HD\,24702. HD\,27989AB is now known to be a BY~Dra star
\citep{2000A+AS..142..275S}. HD\,24702 is constant within the observational
accuracy, whereas HD\,26736 and HD\,26874 scatter slightly more than expected;
see Table~\ref{tab:v1094_std}. For HD\,26874 the effect seems to be random,
whereas HD\,26736 brightened by about 0.02 mag during one observing season.
The light curves are calculated relative to HD\,26874. Observations of
HD\,26736 (except for the bright period) and HD\,24702 (when observed) were
also used, shifting them first to the light level of HD\,26874. 

\begin{table*}
\caption[]{\label{tab:v1094_std}
Photometric data for V1094\,Tau and the comparison stars. For V1094\,Tau, the
$uvby\beta$ information is the mean value outside eclipses. N is the total
number of observations used to form the mean values, and $\sigma$ is the rms
error (per observation) in  mmag. References are: M15 = This paper, O94  =
\cite{1994A+AS..106..257O}. }
\begin{center}
\begin{tabular}{lllrrrrrrrrrrrr} \hline
\hline\noalign{\smallskip}
Object&Sp. Type&Ref.     &$V$&$\sigma$&$b-y$&$\sigma$&$m_1$&$\sigma$&$c_1$ &$\sigma$&N($uvby$)&$\beta$&$\sigma$&N($\beta$)\\
\noalign{\smallskip}
\hline
\noalign{\smallskip}
V1094\,Tau& G0         & M15    & 9.020  & 9& 0.415 & 4& 0.199 & 9& 0.330 &11&206 & 2.596   & 5 &12\\    
\noalign{\smallskip}

 HD\,26736  & G5         & M15    & 8.050  & 8& 0.408 & 5& 0.235 & 8& 0.344 & 5&113 & 2.602   & 5 &13\\ 
          &            & O94    & 8.034  & 5& 0.414 & 3& 0.223 & 4& 0.332 & 6&  1 &         &   &  \\
\noalign{\smallskip}

 HD\,26874  & G4~V       & M15    & 7.824  & 8& 0.432 & 4& 0.270 & 7& 0.315 & 5&171 & 2.596   & 5 &11\\
          &            & O94    &        &  & 0.438 & 3& 0.269 & 4& 0.320 & 6&  1 &         &   &  \\
\noalign{\smallskip}

 HD\,27989AB& G3\,V+G6\,V   & M15    & 7.517  &13& 0.424 & 4& 0.253 & 6& 0.329 & 4& 67 & 2.596   & 6 &16\\
          &            & O94    & 7.532  & 5& 0.424 & 3& 0.246 & 4& 0.326 & 6&  1 &         &   &  \\
\noalign{\smallskip}

 HD\,24702  & G0         & M15    & 7.844  & 5& 0.422 & 5& 0.234 & 9& 0.352 & 5& 48 &         &   &  \\
          &            & O94    &        &  & 0.435 & 3& 0.225 & 4& 0.354 & 6&  1 &         &   &  \\
\hline
\end{tabular}
\end{center}
\end{table*}

 We also obtained  light curves of  V1094~Tau with the WebScope instrument at
the NF/ Observatory (NFO) located near Silver City, New Mexico and the
Undergraduate Research Studies in Astronomy (URSA) WebScope on the roof of the
Kimpel Hall on the University of Arkansas campus at Fayetteville.  The NFO
instrument consists of a 24-inch Cassegrain reflector with a field-widening
correcting lens near the focus housed in a roll-off roof structure
\citep{2008PASP..120..992G}. At the focus is a camera based on the Kodak
KAF-4301E charge coupled device (CCD) with a field of view of about
27$\times$27 arcmin. V1094~Tau was observed at the NFO on 116 nights between 2
Jan 2006 and 24 Mar 2012, producing a total of 5714 observations from 30
second exposures with a Bessel V filter.

 The URSA WebScope uses a 10-inch Meade LX200 Schmidt-Cassegrain telescope
scope with a Santa Barbara Imaging Group (SBIG) ST8 CCD camera, housed in a
Technical Innovations RoboDome enclosure. V1094~Tau was observed on 91 nights
between 6 Mar 2001 and 25 Mar 2012, producing a total of 8085 observations
from 30 second exposures, also with a Bessel V filter.

\begin{figure*}
\mbox{\includegraphics[width=0.98\textwidth]{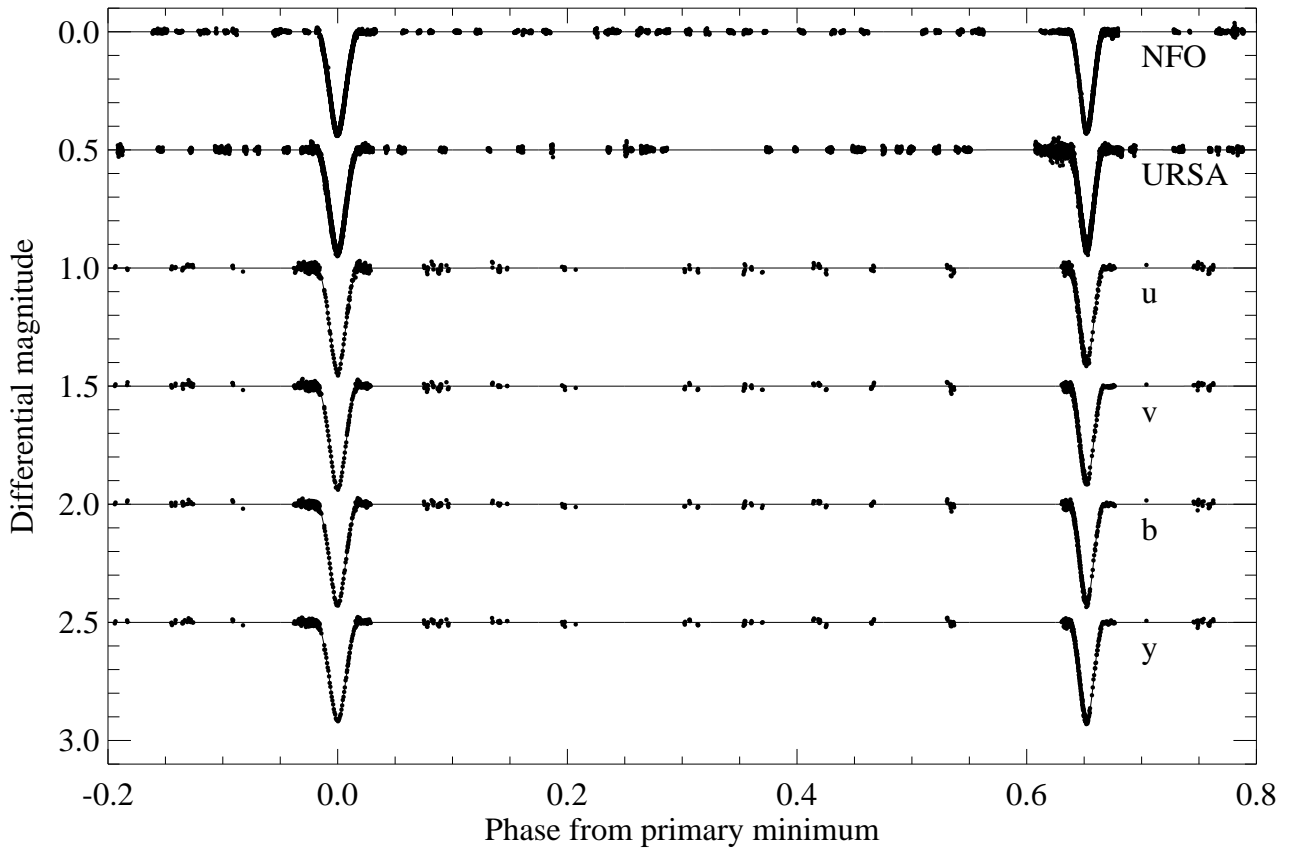}}
\caption{Light curves of V1094~Tau obtained with the SAT (Str\"{o}mgren u, v,
b, and y) and at the  NFO  and URSA observatories (V-band). For clarity,
differential magnitudes are plotted offset in multiples of 0.5 magnitudes.
\label{lcplot}}
\end{figure*}

The images were automatically measured by using the applications {\tt
Multi-Measure} and {\tt Measure} written by author Lacy. The software was used
to locate
the stars of interest in the calibrated images and to perform background
subtraction and aperture photometry in a region 22 arcsec square around each
star in the NFO  images and  30 arcsec square for the URSA images. The
comparison star used was HD~284196 and the check star was HD~284197.
Differences in the atmospheric extinction were corrected for all three stars.
To form the differential magnitudes for analysis we used the flux of both
comparison and check star for the NFO images, whereas in the URSA images only
the comparison star HD 284196 was used. The differences between the comparison
and check star magnitudes averaged 0.008 mag for the URSA images on 88 nights
and 0.006 mag for NFO images on 118 nights. These magnitude differences are
similar to the residuals of the light curve model fits derived from the
analysis below. The complete light curves obtain using the NFO, URSA and SAT
telescopes are shown in Fig.~\ref{lcplot}.

\section{Analysis}

\subsection{Eclipse ephemerides}

We used the {\sc
jktebop}\footnote{\url{www.astro.keele.ac.uk/~jkt/codes/jktebop.html}}
light curve model \citep{2013A&A...557A.119S} to measure times of mid-eclipse
from our SAT, NFO, and URSA photometry. We identified nights of data where our
observations covered the majority of the either the primary or secondary
eclipse. All the parameters of the light curve model except the time of
mid-eclipse and the zero-point of the magnitude scale were fixed at values
similar to those given in Section~\ref{LCFITSECT} for our adopted light curve
solution. We then used a least-squares fit to the data from each night
individually to determine the times of mid-eclipse given. The standard errors
on these times were estimated using the cyclic residual permutation method
(``prayer-bead'' method). 

 We used a least-squares fit of a linear ephemeris to these new times of
mid-eclipse and previously published values to obtain the
optimum values for the reference times of primary and secondary eclipse and
two independent estimates of the orbital period, one from the primary eclipses
and one from the secondary eclipses. If we use the standard errors quoted
on the published times of minimum and from the prayer-bead method for our new
times of minimum we find that the reduced chi-squared value for the fit is
much greater than 1. Our assumption is that this is due to systematic noise
sources that are both instrumental and astrophysical in origin, and that all
the times of minimum are equally affected by these noise sources. In
order to achieve a reduced chi-squared value $\chi^2_r \approx 1$ for the
least-squares fit to the times of primary eclipse we have added 59\,s in
quadrature to the standard error estimates quoted on published values and also
to the estimated standard errors for our new times of minimum. For the times
of secondary eclipse we include an additional 52\,s in the standard error
estimates. The data used and the adopted standard errors are given in
Table~\ref{tminTable}. The resulting linear ephemerides for the heliocentric
Julian Dates (HJD) of primary and secondary eclipse, respectively, are \[ {\rm
HJD} (T_{\rm pri}) = 2454699.33513(19) + 8.9885445(12) E \] and \[{\rm HJD}
(T_{\rm sec}) = 2454498.46151(23) + 8.9885515(13) E, \] where $E$ is the cycle
number and the figures in parentheses are the standard errors in the two final
digits for each free parameter. 
 
 The difference in the orbital period obtained from the times of primary and
secondary eclipse is significant (4-$\sigma$) and is characteristic of binary
stars in which the orientiation of the eccentric orbit is changing (apsidal
motion).

\begin{table}
\caption{Times of mid-eclipse for V1094 Tau. Times are given as UTC heliocentric
Julian date (HJD) and are labelled as primary (``P'') or secondary (``S'')
eclipses. Note that the adopted standard errors used here are not those quoted
in the original sources. \label{tminTable}}
\begin{center}
 \begin{tabular}{@{}llll}
\hline
 \noalign{\smallskip}
HJD-2400000 & & Type & Source \\
 \noalign{\smallskip}
\hline
\hline
 \noalign{\smallskip}
49653.632  &$ \pm$ 0.003  & S & \citet{1995IBVS.4168....1K} \\
49656.762  &$ \pm$ 0.003  & P & \citet{1998IBVS.4544....1K} \\
49680.597  &$ \pm$ 0.003  & S & \citet{1995IBVS.4168....1K} \\
49683.727  &$ \pm$ 0.003  & P & \citet{1998IBVS.4544....1K} \\
49701.7061 &$ \pm$ 0.0014 & P & \citet{1998IBVS.4544....1K} \\
49707.5649 &$ \pm$ 0.0019 & S & \citet{1995IBVS.4168....1K} \\
49710.6950 &$ \pm$ 0.0021 & P & \citet{1998IBVS.4544....1K} \\
49755.6355 &$ \pm$ 0.0021 & P & \citet{1998IBVS.4544....1K} \\
50456.7393 &$ \pm$ 0.0021 & P & \citet{1998IBVS.4544....1K} \\
50474.7156 &$ \pm$ 0.0021 & P & \citet{1998IBVS.4544....1K} \\
52278.2875 &$ \pm$ 0.0011 & S & \citet{2010A+A...509A..18W} \\
52362.3148 &$ \pm$ 0.0011 & P & \citet{2010A+A...509A..18W} \\
52601.87780&$ \pm$ 0.00140& S & V, URSA \\
52628.84325&$ \pm$ 0.00062& S & V, URSA \\
52637.83158&$ \pm$ 0.00065& S & V, URSA \\
52898.4989 &$ \pm$ 0.0011 & S & \citet{2010A+A...509A..18W} \\
52997.3708 &$ \pm$ 0.0028 & S & \citet{2005IBVS.5643....1H} \\
53045.4406 &$ \pm$ 0.0035 & P & \citet{2005IBVS.5657....1H} \\
54438.66519&$ \pm$ 0.00079& P & V, URSA \\
54447.65487&$ \pm$ 0.00098& P & V, NFO \\
54447.65613&$ \pm$ 0.00062& P & u, SAT \\
54447.65604&$ \pm$ 0.00073& P & v, SAT \\
54447.65602&$ \pm$ 0.00093& P & b, SAT \\
54447.65617&$ \pm$ 0.00130& P & y, SAT \\
55157.75013&$ \pm$ 0.00090& P & V, NFO \\
55175.72858&$ \pm$ 0.00062& P & V, NFO \\
55181.59081&$ \pm$ 0.00071& S & V, URSA \\
55202.69394&$ \pm$ 0.00086& P & V, NFO \\
55208.55713&$ \pm$ 0.00069& S & V, URSA \\
55247.63640&$ \pm$ 0.00086& P & V, NFO \\
55274.60204&$ \pm$ 0.00058& P & V, URSA \\
55813.91459&$ \pm$ 0.00101& P & V, URSA \\
55831.89207&$ \pm$ 0.00103& P & V, URSA \\
55831.89275&$ \pm$ 0.00073& P & V, NFO \\
55849.8700 &$ \pm$ 0.0015 & P & \citet{2012IBVS.6011....1D} \\
55855.73235&$ \pm$ 0.00079& S & V, URSA \\
55882.69860&$ \pm$ 0.00057& S & V, NFO \\
55894.81080&$ \pm$ 0.00070& P & V, URSA \\
55936.62917&$ \pm$ 0.00085& S & V, NFO \\
55945.61739&$ \pm$ 0.00078& S & V, NFO \\
55945.62050&$ \pm$ 0.00080& S &  \citet{2012IBVS.6029....1D} \\
 \noalign{\smallskip}
\hline
 \end{tabular}   
\end{center}
\end{table}

 We also analysed the photometry of V1094~Tau described by
\citet{2011MNRAS.416.2477W} but found that these data were too noisy to add
any useful constraints on the ephemeris or the radii of the stars so we do not
discuss them further here. We see no evidence for the periodic variation in
eclipse times claimed by \citet{2010A+A...509A..18W}.

\subsection{Radial velocity measurements}

\begin{table}
\caption{Radial velocities of V1094~Tau from CfA spectroscopy.
\label{RVTable}}
\begin{center}
 \begin{tabular}{@{}rrr}
\hline
 \noalign{\smallskip}
\multicolumn{1}{l}{HJD$-2400000$ }      & Primary [km\,s$^{-1}$]& Secondary
[km\,s$^{-1}$] \\
 \noalign{\smallskip}
\hline
\hline
 \noalign{\smallskip}
    50080.5402 &$ -21.76 \pm 0.46 $&$  30.06  \pm  0.72  $ \\
    50081.5561 &$ -38.87 \pm 0.41 $&$  50.28  \pm  0.64  $ \\
    50090.6749 &$ -41.26 \pm 0.44 $&$  51.80  \pm  0.68  $ \\
    50094.6175 &$  56.08 \pm 0.66 $&$ -54.20  \pm  1.04  $ \\
    50098.5872 &$ -23.76 \pm 0.46 $&$  32.20  \pm  0.72  $ \\
    50108.5424 &$ -39.93 \pm 0.41 $&$  49.52  \pm  0.64  $ \\
    50114.6418 &$  41.45 \pm 0.43 $&$ -38.19  \pm  0.67  $ \\
    50115.4714 &$   7.68 \pm 0.43 $&$  -1.97  \pm  0.67  $ \\
    50118.4844 &$ -46.86 \pm 0.39 $&$  56.77  \pm  0.62  $ \\
    50120.5183 &$  -9.71 \pm 0.41 $&$  18.45  \pm  0.65  $ \\
    50121.5423 &$  52.49 \pm 0.42 $&$ -50.82  \pm  0.66  $ \\
    50126.6304 &$ -41.43 \pm 0.63 $&$  50.12  \pm  0.99  $ \\
    50127.5361 &$ -46.25 \pm 0.49 $&$  58.01  \pm  0.77  $ \\
    50140.5293 &$  83.47 \pm 0.41 $&$ -81.43  \pm  0.65  $ \\
    50143.5077 &$ -22.22 \pm 0.44 $&$  32.63  \pm  0.68  $ \\
    50144.5064 &$ -40.16 \pm 0.40 $&$  51.78  \pm  0.63  $ \\
    50146.5016 &$ -38.69 \pm 0.48 $&$  49.69  \pm  0.75  $ \\
    50147.5844 &$  -4.42 \pm 0.43 $&$  12.59  \pm  0.67  $ \\
    50152.4997 &$ -22.40 \pm 0.45 $&$  33.31  \pm  0.71  $ \\
    50153.5028 &$ -40.55 \pm 0.45 $&$  51.12  \pm  0.70  $ \\
    50154.5311 &$ -46.08 \pm 0.43 $&$  58.03  \pm  0.67  $ \\
    50155.4860 &$ -38.94 \pm 0.43 $&$  48.45  \pm  0.68  $ \\
    50156.4965 &$  -8.02 \pm 0.40 $&$  16.87  \pm  0.63  $ \\
    50170.5041 &$ -22.81 \pm 0.44 $&$  33.28  \pm  0.68  $ \\
    50173.5107 &$ -38.12 \pm 0.38 $&$  48.41  \pm  0.60  $ \\
    50176.5112 &$  83.02 \pm 0.44 $&$ -82.74  \pm  0.69  $ \\
    50179.5248 &$ -24.62 \pm 0.52 $&$  34.14  \pm  0.82  $ \\
    50336.8741 &$  25.92 \pm 0.42 $&$ -22.09  \pm  0.66  $ \\
    50346.8207 &$  81.42 \pm 0.42 $&$ -80.38  \pm  0.66  $ \\
    50348.7661 &$  23.78 \pm 0.45 $&$ -17.67  \pm  0.70  $ \\
    50350.8059 &$ -34.31 \pm 0.68 $&$  44.49  \pm  1.07  $ \\
    50352.8106 &$ -45.10 \pm 0.41 $&$  54.89  \pm  0.65  $ \\
    50356.8289 &$  64.68 \pm 0.42 $&$ -62.93  \pm  0.66  $ \\
    50360.7872 &$ -45.06 \pm 0.44 $&$  55.20  \pm  0.68  $ \\
    50363.8342 &$  26.13 \pm 0.43 $&$ -21.03  \pm  0.68  $ \\
    50443.6832 &$ -26.87 \pm 0.46 $&$  36.25  \pm  0.73  $ \\
    50449.5930 &$ -32.42 \pm 0.52 $&$  41.67  \pm  0.82  $ \\
    50456.5688 &$  26.22 \pm 0.49 $&$ -20.28  \pm  0.77  $ \\
    50460.5754 &$ -45.66 \pm 0.50 $&$  56.81  \pm  0.78  $ \\
    50462.4976 &$  13.02 \pm 0.58 $&$  -5.35  \pm  0.91  $ \\
    50464.6145 &$  67.50 \pm 0.50 $&$ -66.16  \pm  0.79  $ \\
    50466.6979 &$ -13.93 \pm 0.49 $&$  21.60  \pm  0.77  $ \\
    50472.5453 &$  77.67 \pm 0.50 $&$ -76.34  \pm  0.78  $ \\
    50474.6912 &$  20.80 \pm 0.50 $&$ -15.25  \pm  0.79  $ \\
    50477.5770 &$ -44.25 \pm 0.44 $&$  55.52  \pm  0.69  $ \\
    50478.5223 &$ -45.19 \pm 0.49 $&$  56.62  \pm  0.77  $ \\
    50481.6877 &$  81.27 \pm 0.59 $&$ -82.90  \pm  0.93  $ \\
    50493.5115 &$  -8.87 \pm 0.41 $&$  15.61  \pm  0.65  $ \\
    50495.4914 &$ -43.36 \pm 0.39 $&$  55.08  \pm  0.62  $ \\
    50503.5415 &$ -32.98 \pm 0.50 $&$  43.68  \pm  0.78  $ \\
    50505.5246 &$ -45.01 \pm 0.54 $&$  55.96  \pm  0.84  $ \\
    50516.5760 &$  21.87 \pm 0.56 $&$ -17.99  \pm  0.87  $ \\
    50523.5045 &$ -44.86 \pm 0.57 $&$  56.30  \pm  0.89  $ \\
    50527.5177 &$  68.51 \pm 0.57 $&$ -66.85  \pm  0.89  $ \\
    50531.5400 &$ -45.01 \pm 0.54 $&$  54.87  \pm  0.84  $ \\
    50535.4986 &$  77.96 \pm 0.55 $&$ -76.90  \pm  0.86  $ \\
    50536.5040 &$  68.85 \pm 0.54 $&$ -67.56  \pm  0.84  $ \\
    50538.5021 &$ -11.08 \pm 0.57 $&$  17.92  \pm  0.89  $ \\
    50540.5290 &$ -45.28 \pm 0.58 $&$  55.64  \pm  0.91  $ \\
\hline
 \end{tabular}   
\end{center}
\end{table} 

\subsubsection{CfA spectroscopy}

 All our CfA spectra appear double-lined. Radial velocities were obtained
using the two-dimensional cross-correlation technique {\sc todcor}
\citep{1994ApJ...420..806Z}, with templates chosen from a large library of
calculated spectra based on model atmospheres by R.\ L.\ Kurucz
\citep[see][]{1994A+A...287..338N, 2002AJ....124.1144L}. The four main
parameters of the templates are the effective temperature ${\rm T}_{\rm eff}$,
projected equatorial rotational velocity $v \sin i$, metallicity [m/H], and
surface gravity $\log g$. The parameters that have the largest effect on the
measured radial velocities  are ${\rm T}_{\rm eff}$ and $v \sin i$.
Consequently, we held $\log g$ fixed at 4.0 for the hotter and more massive
star (hereafter star 1, or primary star)  and 4.5 for the cooler one (star 2,
secondary star), which are near the final values reported below in
Sect.~\ref{sec:absdim}, and we assumed solar metallicity.  The optimum ${\rm
T}_{\rm eff}$ and $v \sin i$ values were determined by running grids of
cross-correlations, seeking the maximum of the correlation coefficient
averaged over all exposures and weighted by the strength of each spectrum
\citep[see][]{2002AJ....123.1701T}. The projected equatorial rotational
velocities we obtained are $v \sin i = 9.0 \pm 2.0$\,km\,s$^{-1}$ for star 1
and $v \sin i = 4.4 \pm 2.0$\,km\,s$^{-1}$ for star 2. The $v \sin i$ value
for star 2 is much less than the instrumental broadening and so is very
sensitive to the value of the macroturbulence assumed in the model grid. The
effective temperatures derived are ${\rm T}_{\rm eff, 1} = 5860\pm100$\,K for
star 1 and ${\rm T}_{\rm eff, 2} = 5780\pm100$\,K for star 2. The uncertainty
in these values has little effect on the measured radial velocities.

As in previous studies using similar spectroscopic material, we made an
assessment of potential systematic errors in our radial velocities that may
result from residual line blending as well as lines shifting in and out of our
narrow spectral window as a function of orbital phase
\citep[see][]{1996A+A...314..864L}. We did this by performing numerical
simulations analogous to those described by \cite{1997AJ....114.2764T}, and we
applied corrections to the raw velocities based on these simulations to
mitigate the effect. The corrections were typically less than
0.5\,km\,s$^{-1}$ for both stars.

Finally, the stability of the zero-point of our velocity system was
monitored by taking nightly exposures of the dusk and dawn sky, and
small run-to-run corrections (typically under 1\,km\,s$^{-1}$) were applied to
the velocities as described by \cite{1992ASPC...32..110L}. The adopted
heliocentric velocities including all corrections are listed in
Table~\ref{RVTable}.

\subsubsection{INT spectroscopy}

 We also used {\sc todcor} to measure the radial velocities of both stars from
our INT spectra, but for these measurements we used spectra of HD\,216435 (G0V)
and HD\,115617 (G5V) obtained from a library of high-resolution stellar spectra
\citep{2003Msngr.114...10B} as templates for the primary and secondary stars,
respectively. Both the INT spectra and the template spectra were interpolated
onto a uniform logarithmic wavelength grid equivalent to 7.23\,km\,s$^{-1}$ per
pixel using quadratic interpolation. The radial velocities were derived by
fitting a minimum curvature surface to the peak of the two-dimensional cross
correlation function and interpolating to the point of maximum correlation.
The accuracy of the radial velocity measurements that can be obtained from our
INT spectra is limited to about 1\,km\,s$^{-1}$ by instrumental effects
(motion of the star in the slit, flexure, etc.) so we did not attempt to make
any corrections for blending, etc. as we did for the CfA spectra.  Radial
velocities for spectra obtained near primary or secondary eclipse were found
to be unreliable and are not reported here. The radial velocities derived from
the 48 remaining spectra are given in Table~\ref{RVINTTable}.

\begin{table}
\caption{Radial velocity measurements from INT spectroscopy for V1094 Tau. The
standard errors in the primary and secondary radial velocities estimated from
the residuals for a least-squares fit of a Keplerian orbit are
1.4\,km\,s$^{-1}$ and
1.7\,km\,s$^{-1}$, respectively.
\label{RVINTTable}}
\begin{center}
 \begin{tabular}{@{}rrr}
\hline
 \noalign{\smallskip}
\multicolumn{1}{l}{HJD$-2400000$ }        & Primary & Secondary  \\
& [km/s]  & [km/s]     \\
 \noalign{\smallskip}
\hline
\hline
 \noalign{\smallskip}
52561.5866  &$   -28.94 $  & $  32.28$\\
52561.5903  &$   -27.30 $  & $  33.68$\\
52561.5940  &$   -26.60 $  & $  34.94$\\
52561.5977  &$   -29.11 $  & $  32.46$\\
52561.6014  &$   -28.97 $  & $  32.56$\\
52561.7507  &$   -31.40 $  & $  35.83$\\
52561.7544  &$   -36.43 $  & $  31.43$\\
52561.7581  &$   -33.42 $  & $  34.26$\\
52561.7618  &$   -37.48 $  & $  30.59$\\
52561.7655  &$   -37.51 $  & $  30.45$\\
52561.7789  &$   -32.93 $  & $  35.66$\\
52561.7826  &$   -35.62 $  & $  32.86$\\
52562.5627  &$   -43.50 $  & $  49.21$\\
52562.5664  &$   -44.83 $  & $  47.26$\\
52562.5701  &$   -45.01 $  & $  47.22$\\
52562.6743  &$   -44.66 $  & $  50.30$\\
52562.6781  &$   -44.90 $  & $  50.23$\\
52562.6818  &$   -44.38 $  & $  51.03$\\
52562.7685  &$   -46.72 $  & $  50.16$\\
52562.7722  &$   -48.43 $  & $  48.51$\\
52562.7759  &$   -46.51 $  & $  50.33$\\
52563.6158  &$   -49.68 $  & $  53.96$\\
52563.6195  &$   -49.58 $  & $  54.07$\\
52563.6232  &$   -49.51 $  & $  53.58$\\
52563.7631  &$   -48.04 $  & $  54.07$\\
52563.7668  &$   -48.84 $  & $  52.88$\\
52563.7705  &$   -48.88 $  & $  52.98$\\
52566.7185  &$    67.66 $  & $ -71.19$\\
52566.7222  &$    66.88 $  & $ -72.05$\\
52566.7259  &$    66.96 $  & $ -72.84$\\
52566.7297  &$    67.19 $  & $ -72.72$\\
52566.7334  &$    67.76 $  & $ -72.67$\\
52566.7611  &$    69.18 $  & $ -74.22$\\
52566.7648  &$    69.41 $  & $ -74.14$\\
52566.7686  &$    70.05 $  & $ -73.39$\\
52566.7723  &$    70.28 $  & $ -74.14$\\
52566.7760  &$    70.54 $  & $ -74.16$\\
52566.7806  &$    70.42 $  & $ -74.19$\\
52566.7843  &$    69.21 $  & $ -76.27$\\
52566.7867  &$    70.69 $  & $ -77.47$\\
52568.6252  &$    33.26 $  & $ -35.50$\\
52568.6289  &$    33.09 $  & $ -34.80$\\
52568.6325  &$    32.77 $  & $ -34.66$\\
52568.7407  &$    28.53 $  & $ -30.12$\\
52568.7443  &$    27.30 $  & $ -31.13$\\
52568.7480  &$    26.81 $  & $ -31.41$\\
52570.7502  &$   -32.28 $  & $  35.55$\\
52570.7539  &$   -33.54 $  & $  34.33$\\ 
 \noalign{\smallskip}
\hline
 \end{tabular}   
\end{center}
\end{table}

\subsection{Spectroscopic orbit and apsidal motion}

 We first performed least-squares fits of Keplerian orbits to the three sets of
radial velocity data available to us, those from the  CfA spectra, those from
the INT/IDS spectra, and the radial velocities published by
\citet{2003Obs...123..203G}. The results from the three data sets were found
to be consistent with each other except for the differences in the zero-point
of the radial velocity scale between the different data sets.  The offset
between the INT radial velocities and the other radial velocity data is small
compared to the instrumental resolution (14\,km\,s$^{-1}$) and is likely to be
due to the uncertainty in the zero-point of the wavelength calibration for
these spectra. The zero-point of the CfA radial velocity scale is within
0.14\,km\,s$^{-1}$ of the absolute reference frame set by minor planets, which
we have observed regularly with the same instrument for 25 years. 

 There is clear evidence for apsidal motion in V1094~Tau from the measured
times of mid-eclipse. There is also information about the rate of apsidal
motion in the radial velocity data. In order to obtain the best possible
estimate of the apsidal period, $U$, and to ensure that there is no systematic
error in parameters of the spectroscopic orbit due to the variation in the
longitude of periastron, $\omega$, we used a program called {\sc omdot} to
perform a simultaneous least-squares fit to all three radial velocity data
sets and all the measured times of mid-eclipse for the parameters of a
Keplerian orbit in which the longitude of periastron changes at a constant
rate $\dot\omega = 2\pi/U$. The times of mid-eclipse are computed by
calculating the times when the projected separation of the stars is at a
minimum. The optimal solution is obtained using the Levenburg-Marquardt
algorithm \citep{1992nrfa.book.....P}. We were careful to assign accurate
standard errors to all the data so that the relative weighting of the
different data sets is correct and the standard error estimates for the free
parameters are accurate. The free parameters in the fit were: a reference time
of mid-eclipse, ${\rm T}_0$; the anomalistic period, $P_{\rm anom}$; the
orbital eccentricty, $e$; the longitude of periastron at time ${\rm T}_0$,
$\omega_0$; $\dot\omega$; the semi-amplitudes of the spectroscopic orbits,
$K_1$ and $K_2$; the  radial velocity of the binary centre-of-mass for the CfA
data, $\gamma_{\rm CfA}$; two offsets between the different radial velocity
scales, $\gamma_{\rm GB}-\gamma_{\rm CfA}$ and $\gamma_{\rm INT}-\gamma_{\rm
CfA}$. The orbital inclination has a negligible effect on the derived times of
minimum and the spectroscopic orbit so we fix this quantity at the value $i =
88.25^{\circ}$. The sidereal period (mean time between eclipses) is $P_{\rm
sid} = \left(1/P_{\rm anom} + 1/U\right)^{-1}$. The results from our
least-squares fit are given in Table~\ref{tab:v1094_omdot}. 

\begin{figure}
\mbox{\includegraphics[width=0.49\textwidth]{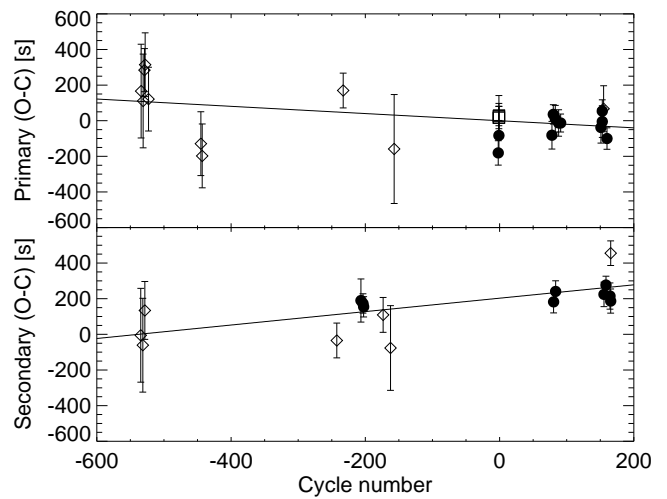}}
\caption{Residuals from a linear ephemeris derived from the values of ${\rm
T}_0$ and $P_{\rm sid}$ given in Table~\ref{tab:v1094_omdot} (points) together
with the fit to these values from {\sc omdot} for an eccentric orbit with
apsidal motion (solid lines). Secondary eclipse is assumed to occur at phase
0.652 for this plot. Published times of mid-eclipse are plotted using diamond
symbols, new times of mid-eclipse are plotted with dots (V-band) or open
squares (SAT). \label{omdotFitTmin}} \end{figure}

\begin{figure}
\mbox{\includegraphics[width=0.49\textwidth]{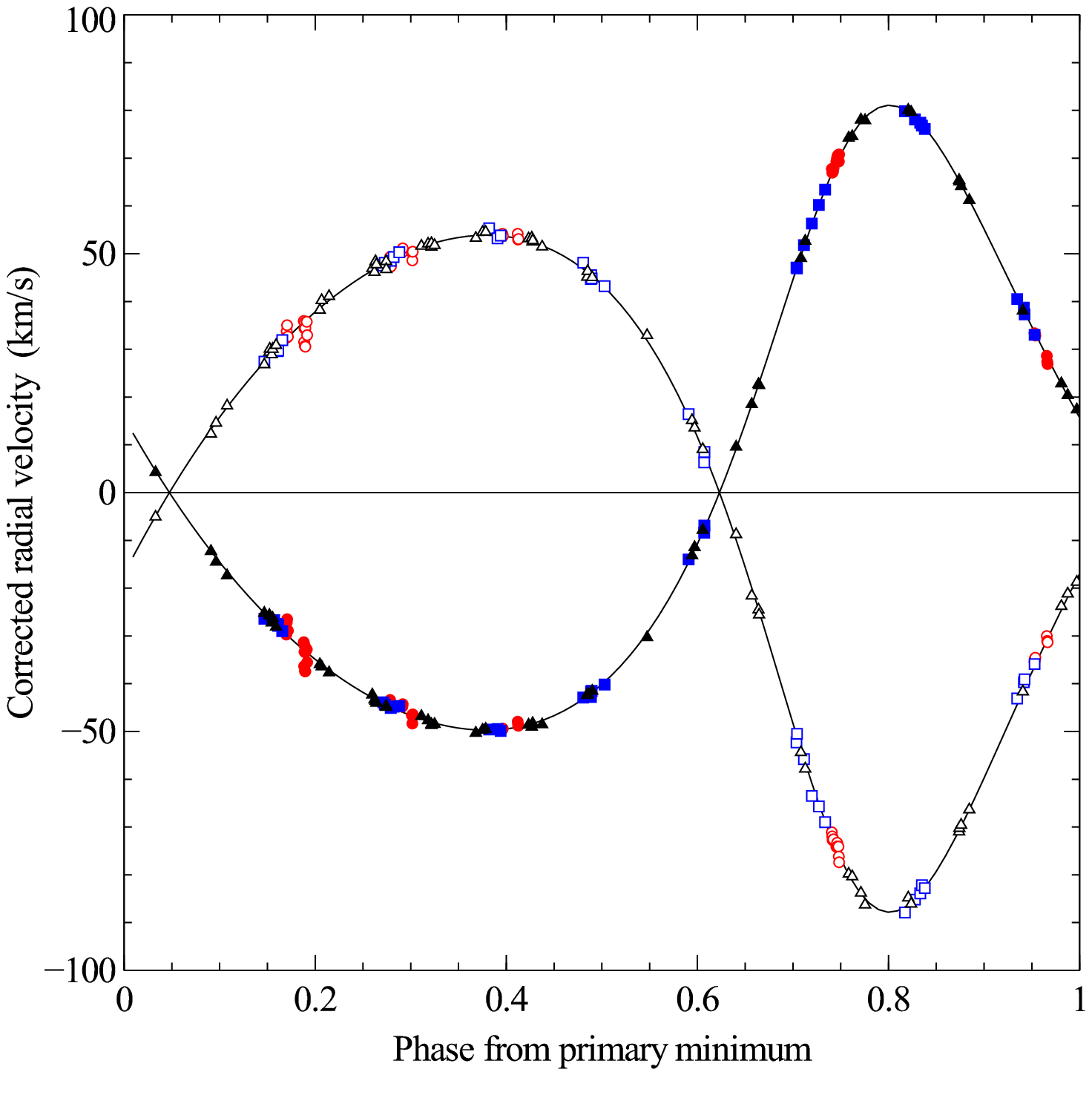}}
\caption{Radial velocities for the primary (filled symbols) and secondary
(open symbols) components of V1094~Tau relative to their
barycentre. Triangles(black) are CfA data, circles (red) are INT data and
squares (blue) are the data from \citet{2003Obs...123..203G}. Spectroscopic
orbits are shown (solid lines) for parameters given in
Table~\ref{tab:v1094_omdot} but with $\omega$ fixed at the value
$333.61^{\circ}$. A colour version of this figure is available in the on-line
version of this article. \label{rvall}} \end{figure}

 The fit to the observed times of mid-eclipse is shown in
Fig.~\ref{omdotFitTmin}. The variation in $\omega$ over the time covered by
radial velocity measurements turns out to be small ($<0.2^{\circ}$) so we show
the fit to the radial velocities for a fixed value of $\omega$ and the
best-fit values of $K_1$ and $K_2$ in Fig.~\ref{rvall}.

\begin{table}
\caption[]{\label{tab:v1094_omdot}
 Parameters of the orbit for V1094~Tau derived using {\sc omdot}.  The number
of eclipse times is $N_{\rm tmin}$, the number of radial velocity
measurements for each star are $N_{\rm rv,1}$ and $N_{\rm rv,2}$ and the
standard deviation of the residuals for each of these data sets are
$\sigma_{\rm tmin}$, $\sigma_{\rm rv,1}$ and $\sigma_{\rm rv,2}$, respectively.
The contribution of the eclipse times and radial velocity  measurements for
each star to the total value of $\chi^2$ for the least-squares fit are
$\chi^2_{\rm tmin}$, $\chi^2_{\rm rv,1}$ and $\chi^2_{\rm rv,2}$, respectively.
Other symbols are defined in the text.
}
\begin{center}
\begin{tabular}{lrl} \hline
\hline\noalign{\smallskip}
Parameter & Value & Notes \\
\noalign{\smallskip}
\hline
${\rm T}_0$   &   2454537.54129 $\pm$ 0.00019 \\
$P_{\rm anom}$ [d]  &  8.9885627  $\pm$ 0.0000036 \\
$e$    & 0.26767 $\pm$ 0.00037 \\
$\omega_0$ [$^{\circ}$] &   333.58 $\pm$  0.17 \\
$\dot{\omega}$ [radians/period] & 0.0000107 $\pm$ 0.0000027 \\
$i$ [$^{\circ}$]  & 88.25 & Fixed.  \\
$K_1$ [km\,s$^{-1}$] &   65.38 $\pm$  0.07 \\
$K_2$ [km\,s$^{-1}$] &   70.83 $\pm$  0.12 \\
$\gamma_{\rm CfA}$  &    3.45 $\pm$   0.05 \\
$\gamma_{\rm GB}-\gamma_{\rm CfA}$&    1.15  $\pm$     0.09  \\
$\gamma_{\rm INT}-\gamma_{\rm CfA}$& $ -3.53  \pm     0.16  $\\
\noalign{\smallskip}
$P_{\rm sid}$   &  8.988547   \\
$U$ [y]         &    $  14500 \pm 3700 $\\
\noalign{\smallskip}
$\sigma_{\rm tmin}$ [d]&  0.00125\\
$N_{\rm tmin}$ &  41\\
$\chi^2_{\rm tmin}$ &     39.2 \\
\noalign{\smallskip}
$\sigma_{\rm rv,1}$ [km\,s$^{-1}$]&  0.93\\
$N_{\rm rv,1}$ &  143\\
$\chi^2_{\rm rv,1}$ &    151.9 \\
\noalign{\smallskip}
$\sigma_{\rm rv,2}$ [km\,s$^{-1}$]&  1.17\\
$N_{\rm rv,2}$ &  143\\
$\chi^2_{\rm rv,2}$ &    146.3 \\
\noalign{\smallskip}
$\chi^2$ &    337.4 \\
\hline
\end{tabular}
\end{center}
\end{table}

\subsubsection{V-band and y-band luminosity ratios \label{sec:lratio}}

 There can be some degeneracy between the ratio of the radii and the
luminosity ratio determined from the light curve for partially eclipsing
binaries, so it is useful to include the luminosity ratio from the
spectroscopy as an additional constraint in the least-squares fit. The
analysis of the CfA spectra using {\sc todcor} yields a luminosity ratio in
the 45\AA\ region around 5190\AA\ of  $\ell_{5190} = 0.534 \pm 0.020$. To
convert this value into a luminosity ratio in the V-band we used the synthetic
stellar spectra from the BaSel 3.1 library \citep{2002A+A...381..524W}. We
compared the flux integrated over the V-band response function
\citep{1990PASP..102.1181B} to the flux integrated over the 45\AA\ region
around 5190\AA\ for a range of effective temperature and metallicity ([Fe/H])
similar to those derived below for the two stars in V1094~Tau. By taking the
ratio of these integrated fluxes and then the ratio of that ratio  for the two
stars we can derive a factor to convert $\ell_{5190}$ to a luminosity ratio in
the V-band, $\ell_{\rm V}$. Applying this correction we obtain $\ell_{\rm V} =
0.557\pm0.025$, where the standard error estimate includes  the additional
uncertainty in this estimate due to the errors in the effective temperatures
of the two stars and  in [Fe/H]. A similar calculation for the Str\"{o}mgren
{\it y}-band gives the result $\ell_{\it y} = 0.533\pm0.025$.  If we were to
extrapolate the luminosity ratio at 5190\AA\ to other wavelengths the results
will be sensitive to systematic errors in the synthetic stellar spectra, so we
have only used the luminosity ratio in the V-band and {\it y}-band using this
method.

\subsubsection{Effective temperature and metallicity\label{sec:teff}}
 The two stars in V1094~Tau are quite similar and so a useful estimate of the
metallicity can be obtained by analysing the Str\"{o}mgremn photometry of this
binary as though it were a single star. The reddening was estimated using the
method of \citet{1988A+A...189..173O} applied to the mean photometric colours
out of eclipse from Table~\ref{tab:v1094_std} and was found to be E$({\it
b}-{\it y}) = 0.019\pm0.007$ magnitudes. The calibration of
\citet{2007A+A...475..519H} applied to the de-reddened Str\"{o}mgren
photometry of V1094~Tau then yields the estimate [Fe/H]$ = -0.09 \pm 0.11$,
where the error includes both the random error in the photometry and the
systematic error in the calibration. \citet{2007A+A...475..519H} also provide
an effective temperature calibration which we can apply to our de-reddened
Str\"{o}mgremn photometry to obtain the value $5680\pm70$\,K. This value will
be a weighted average of the individual effective temperatures for the two
stars in V1094~Tau. The two stars have similar effective temperatures so the
luminosity ratio is almost constant at optical wavelenghs, and so the relative
weights between the two stars will be close to the luminosity ratio 
derived above for the {\it y}-band.
 
 The surface brightness ratio derived from the least-squares fits to the light
curves, $J$, provides a useful constraint on the difference between the
effective temperatures of the two stars, $\Delta {\rm T}_{\rm eff} = {\rm
T}_{\rm eff,1} - {\rm T}_{\rm eff,2} $. We used the synthetic stellar spectra
from \citet{1993KurCD..13.....K} to establish calibrations between surface
brightness against T$_{\rm eff}$ for the V-band, {\it y}-band and {\it b}-band
assuming either $\log g = 4.0$ (secondary) or $\log g = 4.5$ (primary). As
expected, the value of $\log g$ has little effect on the predicted surface
brightness. We then interpolated between these calibrations to find the value
of ${\rm T}_{\rm eff,2} $ that gives the observed surface brightness ratio
assuming ${\rm T}_{\rm eff,1} = 5860\pm 100$\,K. We investigated how these
calibrations are affected by the assumed metallicity using the synthetic
stellar spectral library by \citet{2002A+A...381..524W} and found that the
estimate of $\Delta {\rm T}_{\rm eff}$ changes by less than 10\,K if the
assumed value of [Fe/H] is changed by $\pm0.1$\,dex. There is very good
agreement between the $\Delta {\rm T}_{\rm eff}$ values derived using the two
different spectral libraries. The values of  $\Delta {\rm T}_{\rm eff}$
derived by this method are $125 \pm 14$ for the NFO  V-band, $159 \pm 15$ for
the URSA V-band, $170 \pm 14$ for the  {\it y}-band, and $152 \pm 14$ for the 
SAT {\it b}-band. The weighted mean of these estimates for the temperature
difference is $\Delta {\rm T}_{\rm eff} =  150 \pm 10$\,K, where we have
quoted the standard error in the weighted mean based on the scatter between
the four input values. This is only slightly larger (by about 30\%) than the
standard error in the weighted mean calculated from the standard errors of
these values. This shows that the level of systematic error in our $J$ values
is low, i.e., the quoted errors on $J$ are close to the true error in these
values. The Str\"{o}mgren {\it u} and {\it v} bands are strongly affected by
details of the stellar models such as line blanketing and convection, so we
have not attempted the same calculation at these wavelengths.

\begin{figure}
\mbox{\includegraphics[width=0.49\textwidth]{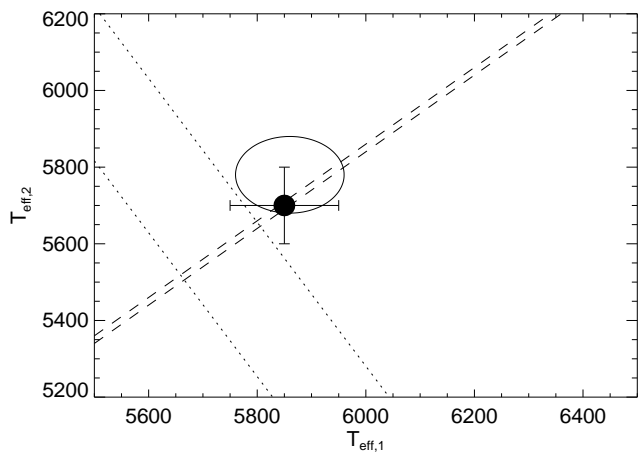}}
\caption{Adopted effective temperatures  and standard errors (points with error
bars) assuming $[{\rm Fe/H}]=-0.09$. Constraints (1-$\sigma$) on the effective
temperatures are indicated as follows: solid lines --  CfA spectroscopy, dotted
lines -- Str\"{o}mgren photometry, dashed lines -- V-band luminosity ratio.
\label{tbest}}
\end{figure}

 The three constraints on the effective temperatures of the stars are shown in
Fig.~\ref{tbest} together with our  adopted values   ${\rm T}_{\rm eff,1} =
5850$\,K and ${\rm T}_{\rm eff,2} = 5700$\,K. We have assumed in this plot
that the value of ${\rm T}_{\rm eff}$  derived from the Str\"{o}mgren
photometry is a weighted average of the two individual ${\rm T}_{\rm eff}$
values with a ratio of weights equal to the luminosity ratio in the {\it
y}-band. Given the level of agreement between the three constraints and the
estimated precision of the CfA effective temperature estimates, we have
adopted estimated standard errors on both these values of $\pm 100$\,K.

\subsubsection{Light curve analysis \label{LCFITSECT}}

 We used {\sc jktebop} version 25 (\citealt{2013A&A...557A.119S} and
references therein) to analyse our six independent light curves of V1094~Tau
({\it uvby}, V$_{\rm NFO}$ and V$_{\rm URSA}$) using the {\sc ebop} light curve
model \citep{1981psbs.conf..111E,1981AJ.....86..102P}. We only included data
from nights covering the primary or secondary eclipse and we modified  {\sc
jktebop} to include offsets in the zero-point of the magnitude scale between
different nights as free parameters in the least-squares fit. This increases
the number of free parameters in the least-squares fit, but it enables us to
check that none of the parameters of interest is strongly affected by
night-to-night offsets in the photometry. For light curves  such as those
presented here where data from different nights overlap in phase we find that
the offsets are well constrained and uncorrelated with other parameters so
including them in the least-squares fits does not cause any problems. 
The offsets typically have values of a few milli-magnitudes with standard
errors of about 1 milli-magnitude, though a few nights have offsets of about
10 milli-magnitudes. Other free parameters in the least-squares fit were: a
normalisation constant, the surface brightness ratio $J = S_2/S_1$, where
$S_1$ is the surface brightness of star 1 at the centre of the stellar disc
and similarly for $S_2$; the sum of the radii relative to semi-major axis,
$(R_{\rm 1}+R_{\rm 2})/a$; the ratio of the radii, $k=R_{\rm 2}/R_{\rm 1}$;
the orbital inclination, $i$;  the phase of primary eclipse, $\Delta$. We
fixed the time of primary eclipse, the orbital period, $e$, and $\omega$ by
using the results from {\sc omdot} described above to calculate these value of
the quantities at the mid-point of the observed data. Separate light curve
solutions that included $e\cos(\omega)$ and $e\sin(\omega)$ as free parameters
showed that the values derived are consistent with those derived using {\sc
omdot}. The variation in these quantities during the span of the observations
due to apsidal motion has a negligible effect on the results. For the V-band
and {\it y}-band light curves the results presented here  include the
luminosity ratio calculated in section \ref{sec:lratio} as an additional
constraint in the least-squares fit. We also tried least-squares fits without
including the luminosity ratio as a constraint and found that this has a
negligible effect on the parameters derived. 

\begin{figure*}
\mbox{\includegraphics[width=0.99\textwidth]{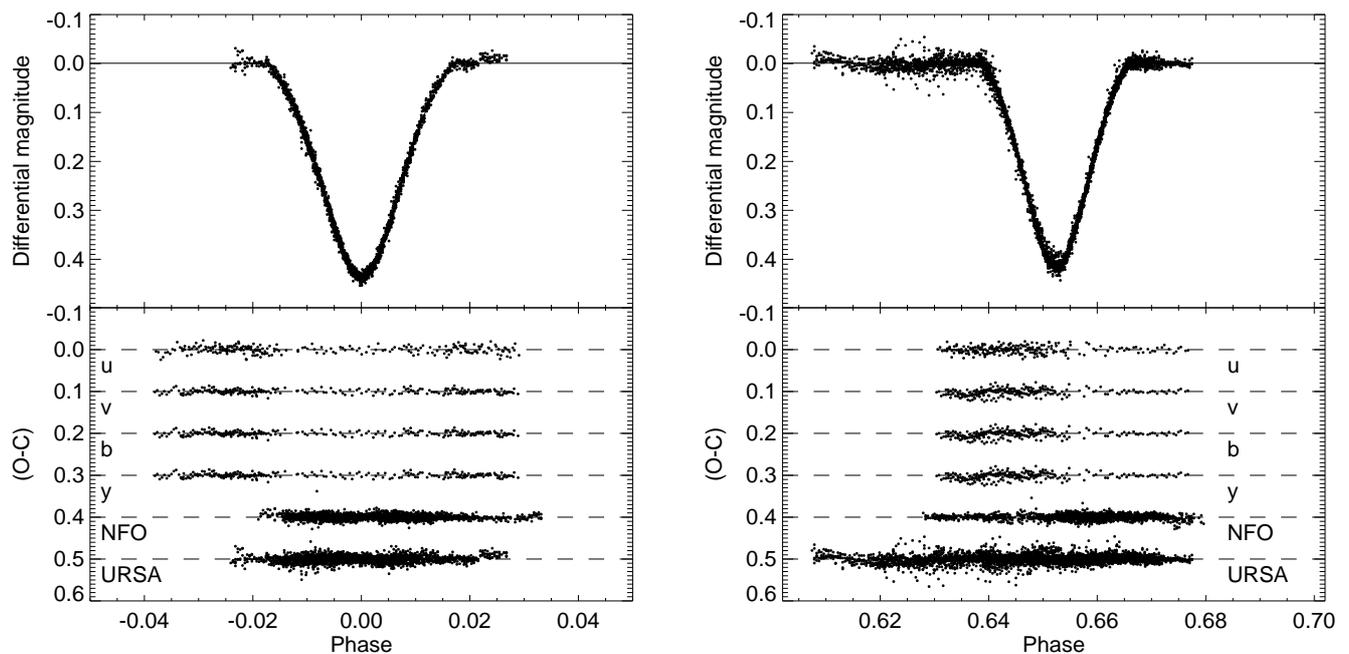}}
\caption{Upper panels: Detail of the {\sc jktebop} model fit to the URSA V-band
light curve of V1094~Tau around  primary eclipse and secondary eclipse. Lower
panels: Residuals from the {\sc jktebop} model fit to our light curves of
V1094~Tau, as labelled, in the region of the primary and secondary eclipse.
For clarity, the residuals are offset in multiples of 0.1 magnitudes. Note
that night-to-night offsets calculated as part of the least-squares fits  have
been applied to the observed magnitudes.
\label{lcfit}}
\end{figure*}

 Gravity darkening coefficients, $\beta$, were taken from the tabulation of
\citet{2011A+A...529A..75C}, although gravity darkening has a negligible
impact on the light curves for these nearly-spherical stars.   We tried a
quadratic limb darkening law  for the fit to the {\it b}-band light curve and
found that there was no improvement in the quality of the fit compared to
a linear limb darkening law and that the parameters of interest are negligibly
affected, so we adopt linear limb darkening laws for all our light curves.
The linear limb darkening coefficients for the two stars used in our analysis
($x_1$ and $x_2$) are given in Table~\ref{tab:v1094_jktebop}. These are
representative of the values obtained from several tabulations of
limb-darkening coefficients for the appropriate effective temperature, gravity
and metallicity of the stars \citep{1993AJ....106.2096V,
1995A+AS..110..329D,2000A+A...363.1081C, 2003A+A...412..241C}. We estimate
that the typical  uncertainties on these values are about 0.04. The
uncertainty on the limb-darkening coefficients has a small effect on the
best-fit values of $R_1/a$ and $R_2/a$ derived from each light curve
($\approx 0.5$\%). We assume that these errors are random so that the effect
on the weighted mean values of $R_1/a$ and $R_2/a$ derived from the six light
curves is negligible. The optimum values of the free parameters and their
standard errors are given in Table~\ref{tab:v1094_jktebop}, where we also
provide the weighted mean values and their standard errors for the
wavelength-independent parameters. The  standard errors quoted are the
standard error in the weighted mean based on the scatter between the six input
values. This is only slightly larger (by about 30\%) than the standard error
in the weighted mean calculated from the standard errors on the six input
values. This shows that the level of systematic error in these free parameters
is low, i.e., the quoted errors on the model parameters are close to the true
error on these values. In particular, we checked that there is no trend in the
estimated inclination versus wavelength that might suggest contamination of
the light curves by third light with a different spectrum to the stars in
V1094~Tau. The best fit to the V-band photometry from URSA around the eclipses
is shown in Fig.~\ref{lcfit}, together with the residuals from the
least-squares fits for all six light curves. 

\begin{table*}
\caption[]{\label{tab:v1094_jktebop}
Light curve parameters for V1094~Tau derived using {\sc jktebop}. The number of
observations used in the fit is $N$ and the standard deviation of the residuals
is $\sigma$. Other symbols are defined in the text. Values preceeded by ``=''
are fixed parameters or constraints in the least-squares fit.}
\begin{center}
\begin{tabular}{@{}lrrrrrrr} \hline
\hline\noalign{\smallskip}
&\multicolumn{1}{c}{\it u}&\multicolumn{1}{c}{\it v}&\multicolumn{1}{c}{\it
b}&\multicolumn{1}{c}{\it y}&\multicolumn{1}{c}{NFO}
  &\multicolumn{1}{c}{URSA}&\multicolumn{1}{l}{Adopted}\\
\hline\noalign{\smallskip}
$(R_1+R_2)/a$            & 0.10804&  0.10830 &  0.10870 &  0.10870 & 0.10737 & 0.10772 &  0.10805 \\
\multicolumn{1}{r}{$\pm$}& 0.00068&  0.00043 &  0.00047 &  0.00035 & 0.00028 & 0.00058 &  0.00026 \\
\noalign{\smallskip}                                                                          
$k=R_2/R_1$              &   0.751&    0.783 &    0.799 &    0.791 &   0.788 &   0.809 &    0.788 \\
\multicolumn{1}{r}{$\pm$}&   0.018&    0.014 &    0.017 &    0.010 &   0.012 &   0.018 &    0.006 \\
\noalign{\smallskip}                                                                          
$R_1/a$                  & 0.06169&  0.06075 &  0.06044 &  0.06069 & 0.06005 & 0.05954 &  0.06050 \\
\multicolumn{1}{r}{$\pm$}& 0.00051&  0.00036 &  0.00043 &  0.00028 & 0.00032 & 0.00048 &  0.00024 \\
\noalign{\smallskip}                                                                          
$R_2/a$                  & 0.04636&  0.04755 &  0.04627 &  0.04801 & 0.04732 & 0.04817 &  0.04744 \\
\multicolumn{1}{r}{$\pm$}& 0.00081&  0.00060 &  0.00071 &  0.00044 & 0.00047 & 0.00076 &  0.00029 \\
\noalign{\smallskip}                                                                          
$e \cos \omega$          & 0.23969&  0.23956 &  0.23957 &  0.23962 & 0.23985 & 0.24000 &          \\
\multicolumn{1}{r}{$\pm$}& 0.00012&  0.00007 &  0.00007 &  0.00006 & 0.00004 & 0.00010 &          \\
\noalign{\smallskip}                                                                                   
$e \sin \omega$          &$-0.11913$&$-0.11915$&$-0.11916$&$-0.11924$&$-0.11886$&$-0.12079$&           \\
\multicolumn{1}{r}{$\pm$}& 0.00003 & 0.00004 & 0.00004 & 0.00006 &  0.00060 &  0.00095 &           \\
\noalign{\smallskip}                                                                             
$i$~[$^{\circ}$]         &    88.31&   88.20&  88.20 & 88.19 &    88.22 &    88.19 & 88.210 \\
\multicolumn{1}{r}{$\pm$}&     0.05&    0.03&   0.04 &  0.03 &     0.03 &     0.04 &  0.014\\
\noalign{\smallskip}                                                                     
$J$                      &    0.851&   0.850&  0.876 & 0.899 &    0.897 &    0.873 &           \\
\multicolumn{1}{r}{$\pm$}&    0.014&   0.008&  0.009 & 0.007 &    0.006 &    0.009 &           \\
\noalign{\smallskip}                                                                       
$\ell=L_2/L_1$           &    0.476&   0.517&  0.554 & 0.559 &    0.553 &    0.567 &           \\
\multicolumn{1}{r}{$\pm$}&    0.016&   0.014&  0.019 & 0.011 &    0.013 &    0.020 &           \\
\noalign{\smallskip}                                                                                   
$x_1$                    &$ =0.821 $&$ =0.810 $&$ =0.753 $&$ =0.662 $&$ =0.662 $&$ =0.662 $&           \\
$x_2$                    &$ =0.841 $&$ =0.822 $&$ =0.769 $&$ =0.679 $&$ =0.678 $&$ =0.678 $&           \\
\noalign{\smallskip}                                                                                   
$N$                      &      492 &      492 &      492 &      492 &     3503 &     5714 &           \\ 
$\sigma$ [mmag]          & 8.5&  5.4&  5.3&  5.2&      6.5 &     10.0 &           \\
\\
\multicolumn{8}{@{}l}{Constraints} \\
$e \cos \omega$          &$ =0.23970$&$=0.23970$&$=0.23970$&$=0.23970$&$=0.23976$&$=0.23964$&           \\
\multicolumn{1}{r}{$\pm$}&  0.00031 &  0.00031 &  0.00031 &  0.00031 &  0.00045 &  0.00045 &           \\
\noalign{\smallskip}
$e\sin\omega$&        $=-0.11909$ &$=-0.11909$&$=-0.11909$&$=-0.11909$&$=-0.11898$&$=-0.11923$& \\
\multicolumn{1}{r}{$\pm$}&  0.00039 &  0.00039 &  0.00039 &  0.00039 & 0.00076 &  0.00076 &  \\
\noalign{\smallskip}
$\ell=L_2/L_1$           &          &          &          &   =0.533 &=0.557 &   =0.557 &           \\
\multicolumn{1}{r}{$\pm$}&          &          &          &    0.025 & 0.025 &    0.025 &           \\
\noalign{\smallskip}
\hline
\end{tabular}
\end{center}
\end{table*}

\subsection{Masses and radii\label{sec:absdim}}
 We have used {\sc
jktabsdim}\footnote{\url{www.astro.keele.ac.uk/~jkt/codes/jktabsdim.html}} to
combine the parameters of the spectroscopic orbit from
Table~\ref{tab:v1094_omdot} and the weighted mean values of $R_1/a$, $R_2/a$
and $i$ from Table~\ref{tab:v1094_jktebop} to derive the masses and radii of
the stars in V1094~Tau with their estimated standard errors given in
Table~\ref{tab:v1094_abspar}.  Table~\ref{tab:v1094_abspar} also provides
estimates of the luminosity and absolute V-band magnitude of the stars based
on the estimates of the stars' effective temperatures in
section~\ref{sec:teff}. The comparison with stellar models discussed below
suggests that V1094~Tau may be slightly more metal-rich than assumed in
section~\ref{sec:teff}. The effective temperatures of the stars estimated from
the CfA spectroscopy also increases by about 70\,K if the assumed values of
[Fe/H] is increased by 0.1\,dex.  For that reason, we also provide in
Table~\ref{tab:v1094_abspar} estimates of the effective temperature,
luminosity and absolute V-band magnitude of the stars for an assumed
metallicity [Fe/H]=+0.14. In both cases, we also provide an estimate of the
distance to V1094~Tau based on the surface-brightness -- effective temperature
relation in the K-band from \citet{2004A+A...426..297K}, and the apparent
K$_s$-band magnitude of V1094~Tau from 2MASS transformed to the Johnson system
\citep[$K=7.468 \pm  0.021$, ][]{2006AJ....131.1163S, 2005ARA+A..43..293B}.

\begin{table}
\caption[]{\label{tab:v1094_abspar}
 Absolute astrophysical parameters for both components of V1094~Tau determined
with {\sc jktabsdim}. Absolute V
magnitudes use bolometric corrections from \citet{1998A+A...333..231B}.
}
\begin{center}
\begin{tabular}{@{}lrr} \hline
\hline\noalign{\smallskip}
 & \multicolumn{1}{l}{Primary} & \multicolumn{1}{l}{Secondary} \\
\noalign{\smallskip}
\hline
\noalign{\smallskip}
Sidereal period [d]  &  \multicolumn{2}{c}{8.9885474 $\pm$ 0.000004} \\
Mass ratio &  \multicolumn{2}{c}{0.9231 $\pm$ 0.0019} \\
Eccentricity & \multicolumn{2}{c}{$0.26755 \pm 0.00004$ } \\
\noalign{\smallskip}
Mass   [$M_{\sun}$] & 1.0965 $\pm$ 0.0040 & 1.0121 $\pm$ 0.0028 \\
Radius [$R_{\sun}$] & 1.4109 $\pm$ 0.0058 & 1.1063 $\pm$ 0.0066 \\
$\log g$ (cgs)   & 4.179 $\pm$ 0.004 & 4.355 $\pm$ 0.005 \\
\noalign{\smallskip}
\multicolumn{3}{@{}l}{Assuming [Fe/H] = $-0.09$}\\
T$_{\rm eff}$ [K] & 5850 $\pm$ 100 &  5700 $\pm$ 100 \\
$\log(L/L_{\sun})$   & 0.32 $\pm$ 0.07 & 0.05 $\pm$ 0.03 \\
M$_{\rm V}$& 4.00 $\pm$ 0.09 & 4.66 $\pm$ 0.08 \\
Distance      [pc]  &  \multicolumn{2}{c}{122   $\pm$ 2  } \\
\noalign{\smallskip}
\multicolumn{3}{@{}l}{Assuming [Fe/H] = $+0.14$}\\
T$_{\rm eff}$ [K] & 5950 $\pm$ 100 &  5800 $\pm$ 100 \\
$\log(L/L_{\sun})$  & 0.35 $\pm$ 0.03 & 0.10 $\pm$ 0.03\\
M$_{\rm V}$& 3.91 $\pm$ 0.09 &  4.57 $\pm$ 0.09 \\
Distance      [pc]  &  \multicolumn{2}{c}{123   $\pm$ 2  } \\
\hline
\end{tabular}
\end{center}
\end{table}

\section{Discussion}

\begin{figure*}
\mbox{\includegraphics[width=0.49\textwidth]{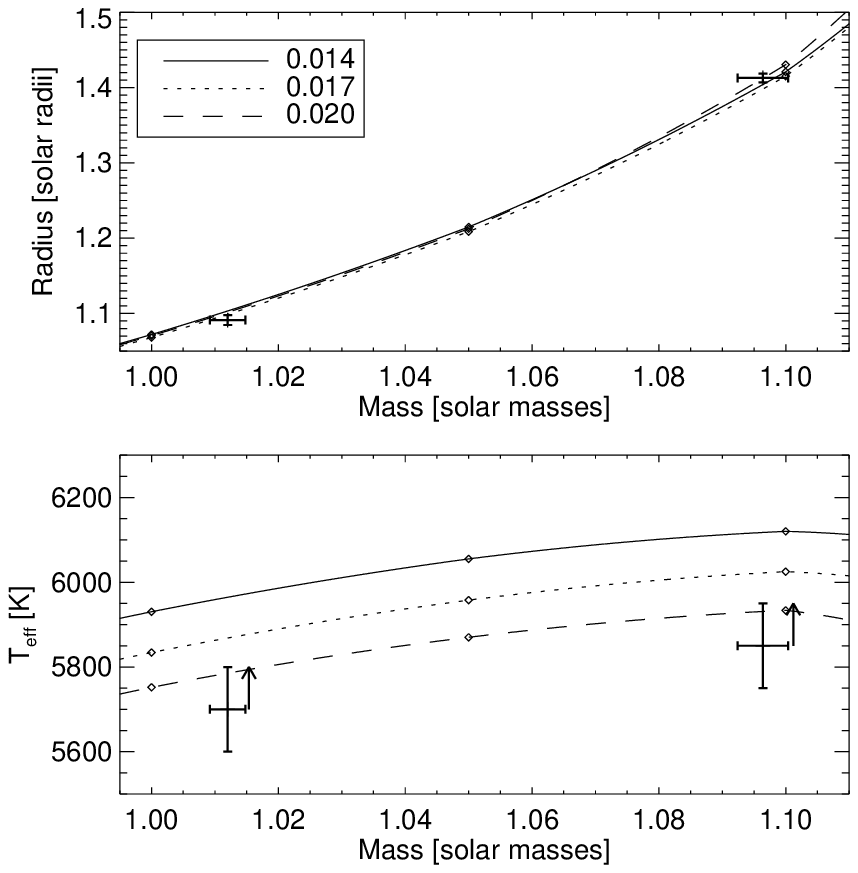}}
\mbox{\includegraphics[width=0.49\textwidth]{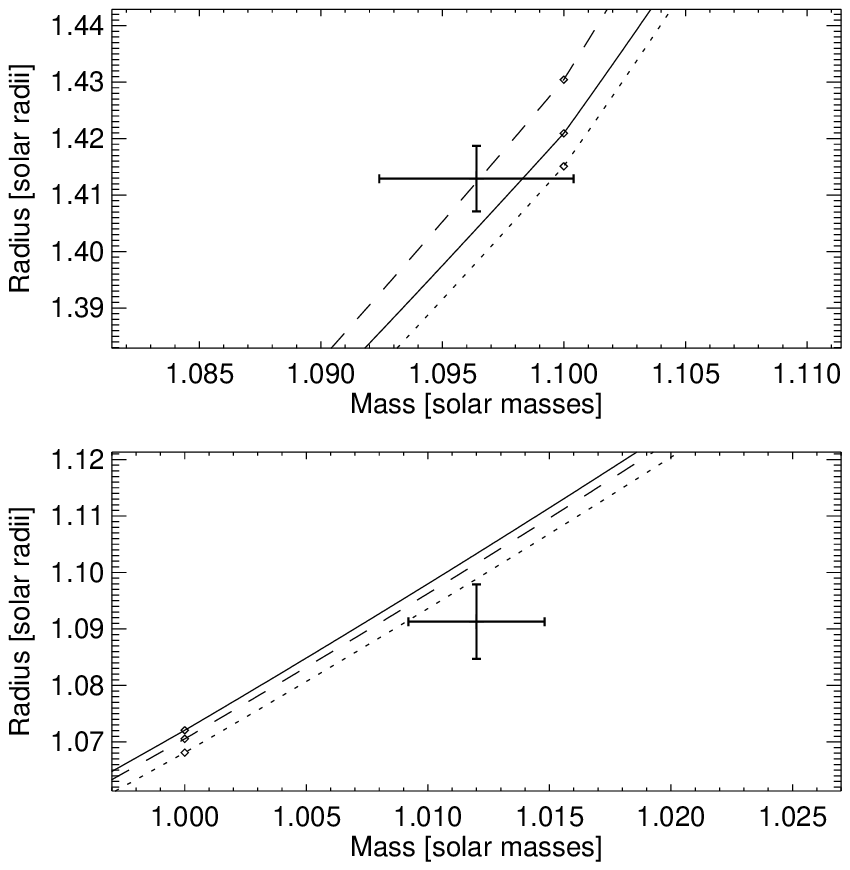}}
\caption{Mass, radius and effective temperature of the two components of
V1094~Tau (errror bars) compared to models from \citet{2012MNRAS.427..127B} for various
values of the initial metal abundance, $Z$. The assumed ages are 5.75\,Gyr,
6.2\,Gyr and 6.6\,Gyr for $Z = 0.014, 0.017, 0.020$, respectively. The arrow in
the  lower-left panel indicates the effect of changing the assumed metallicity from
[Fe/H]$=-0.09$ to [Fe/H]$=+0.14$.  Small diamonds show the model grid
points that have been interpolated to produce the isochrones in these plots.
\label{modelfit}}
\end{figure*}

 Fig.~\ref{modelfit} shows the two stars in V1094~Tau in the mass--radius and
mass--T$_{\rm eff}$ planes compared to various models from
\citet{2012MNRAS.427..127B}.  The primary star is close to the end of its
main-sequence lifetime and so models for this star are sensitive to the
assumed age of the binary. For each value of the metal abundance shown we
have adjusted the age of the models in order to find a good match to the
radius of the more massive star.  For a fixed metallicity we find that we
can determine the age of the system with a precision of about 0.1\,Gyr. If we
assume, as Bressan et al. do,  that the metal abundance of the Sun is
$Z_{\sun} = 0.01774$, and that the helium abundance of V1094~Tau is similar to
the Sun,  then our estimate [Fe/H]$=-0.09\pm0.11$ corresponds to
$Z=0.014\pm0.005$. It is clear that for $Z=0.014$ the effective temperatures
of both stars are too low compared to the models. Increasing the assumed metal
abunance to $Z=0.020$ provides a much better fit to the effective
temperatures, particularly when we account for the increase in our estimates
of T$_{\rm eff}$ if the assumed metallicity is increased. A reasonable fit to
all the observations can be obtained by assuming an intermediate value for the
metal abundance, in which case the age of V1094~Tau is estimated to be about
6\,Gyr. Very similar results are found using other stellar models grids such
as \citet{2006ApJS..162..375V} or \citet{2008ApJS..178...89D}.

 There is good agreement between the projected equatorial velocity of the
primary star quoted by \citet{2003Obs...123..203G} ($9\pm1$\,km\,s$^{-1}$) and
the value we have derived from our CfA spectroscopy ($9.0 \pm
2.0$\,km\,s$^{-1}$). The agreement is less good for the secondary star
($7\pm1$\,km\,s$^{-1}$ versus $4.4 \pm 2.0$\,km\,s$^{-1}$) but the lower value
derived from the CfA spectroscopy is very sensitive to the assumed
macroturbulence parameter used in the stellar model grid because the
rotational broadening in less than the resolution of the instrument
(8.5\,km\,s$^{-1}$). Nevertheless, both the primary and secondary stars
appear to rotate slightly below the rate expected for pseudo-synchronous
rotation ($11.42 \pm 0.05$\,km\,s$^{-1}$ and $8.96 \pm 0.05$\,km\,s$^{-1}$,
respectively). This suggests that V1094~Tau is close to the limit at which
tidal interactions between the stars force them to rotate
pseudo-synchronously with the orbital motion.

\section{Conclusion}
 We have measured the masses of the stars in V1094~Tau to better than 
0.4\% and the radii of these stars to better than 0.6\%. The error
estimates on these values are robust as they are based on the analysis of
multiple high-quality independent data sets. This level of precision and
accuracy is among the best currently available for any solar-type star apart
from the Sun itself. The sample of 95 well-studied eclipsing binary
stars compiled by \citet{2010A+ARv..18...67T} contains only one star with both
its mass and radius measured to this accuracy (HD~124784~B). 

 We find that the current generation of stellar models are able to match the
observed masses and radii of these stars within the constraints currently
available on the effective temperatures of the stars and their metallicity.
These models suggest that the age of V1094~Tau is about 6\,Gyr. The rotation
velocities of the stars suggest that V1094~Tau is close to the limit at which
tidal interactions between the stars force them to rotate pseudo-synchronously
with the orbit.
 
\begin{acknowledgements}

We thank the anonymous referee for their careful consideration of our
manuscript and their constructive comments that have helped to improve the
paper. Thanks to Dr. A.~W. Neely for the maintenance and operation of the NFO
and for preliminary processing and distribution of the images. GT acknowledges
partial support from NSF grant AST-1007992. JVC participated fully in the data
collection and analysis up to the time of his death, but bears no
responsibility for the final text of this paper.
\end{acknowledgements}

\bibliographystyle{aa} 
\bibliography{wasp}

\begin{thebibliography}{46}
\expandafter\ifx\csname natexlab\endcsname\relax\def\natexlab#1{#1}\fi

\bibitem[{{Bagnulo} {et~al.}(2003){Bagnulo}, {Jehin}, {Ledoux}, {Cabanac},
  {Melo}, {Gilmozzi}, \& {The ESO Paranal Science Operations
  Team}}]{2003Msngr.114...10B}
{Bagnulo}, S., {Jehin}, E., {Ledoux}, C., {et~al.} 2003, The Messenger, 114, 10

\bibitem[{{Bessell}(1990)}]{1990PASP..102.1181B}
{Bessell}, M.~S. 1990, \pasp, 102, 1181

\bibitem[{{Bessell}(2005)}]{2005ARA+A..43..293B}
{Bessell}, M.~S. 2005, \araa, 43, 293

\bibitem[{{Bessell} {et~al.}(1998){Bessell}, {Castelli}, \&
  {Plez}}]{1998A+A...333..231B}
{Bessell}, M.~S., {Castelli}, F., \& {Plez}, B. 1998, \aap, 333, 231

\bibitem[{{Bressan} {et~al.}(2012){Bressan}, {Marigo}, {Girardi}, {Salasnich},
  {Dal Cero}, {Rubele}, \& {Nanni}}]{2012MNRAS.427..127B}
{Bressan}, A., {Marigo}, P., {Girardi}, L., {et~al.} 2012, \mnras, 427, 127

\bibitem[{{Claret}(2000)}]{2000A+A...363.1081C}
{Claret}, A. 2000, \aap, 363, 1081

\bibitem[{{Claret} \& {Bloemen}(2011)}]{2011A+A...529A..75C}
{Claret}, A. \& {Bloemen}, S. 2011, \aap, 529, A75

\bibitem[{{Claret} \& {Hauschildt}(2003)}]{2003A+A...412..241C}
{Claret}, A. \& {Hauschildt}, P.~H. 2003, \aap, 412, 241

\bibitem[{{Diaz-Cordoves} {et~al.}(1995){Diaz-Cordoves}, {Claret}, \&
  {Gimenez}}]{1995A+AS..110..329D}
{Diaz-Cordoves}, J., {Claret}, A., \& {Gimenez}, A. 1995, \aaps, 110, 329

\bibitem[{{Diethelm}(2012{\natexlab{a}})}]{2012IBVS.6011....1D}
{Diethelm}, R. 2012{\natexlab{a}}, Information Bulletin on Variable Stars,
  6011, 1

\bibitem[{{Diethelm}(2012{\natexlab{b}})}]{2012IBVS.6029....1D}
{Diethelm}, R. 2012{\natexlab{b}}, Information Bulletin on Variable Stars,
  6029, 1

\bibitem[{{Dotter} {et~al.}(2008){Dotter}, {Chaboyer}, {Jevremovi{\'c}},
  {Kostov}, {Baron}, \& {Ferguson}}]{2008ApJS..178...89D}
{Dotter}, A., {Chaboyer}, B., {Jevremovi{\'c}}, D., {et~al.} 2008, \apjs, 178,
  89

\bibitem[{{Enoch} {et~al.}(2010){Enoch}, {Collier Cameron}, {Parley}, \&
  {Hebb}}]{2010A+A...516A..33E}
{Enoch}, B., {Collier Cameron}, A., {Parley}, N.~R., \& {Hebb}, L. 2010, \aap,
  516, A33+

\bibitem[{{Etzel}(1981)}]{1981psbs.conf..111E}
{Etzel}, P.~B. 1981, in Photometric and Spectroscopic Binary Systems, ed.
  {E.~B.~Carling \& Z.~Kopal}, 111

\bibitem[{{Grauer} {et~al.}(2008){Grauer}, {Neely}, \&
  {Lacy}}]{2008PASP..120..992G}
{Grauer}, A.~D., {Neely}, A.~W., \& {Lacy}, C.~H.~S. 2008, \pasp, 120, 992

\bibitem[{{Griffin} \& {Boffin}(2003)}]{2003Obs...123..203G}
{Griffin}, R.~F. \& {Boffin}, H.~M.~J. 2003, The Observatory, 123, 203

\bibitem[{{Holmberg} {et~al.}(2007){Holmberg}, {Nordstr{\"o}m}, \&
  {Andersen}}]{2007A+A...475..519H}
{Holmberg}, J., {Nordstr{\"o}m}, B., \& {Andersen}, J. 2007, \aap, 475, 519

\bibitem[{{Horne}(1986)}]{1986PASP...98..609H}
{Horne}, K. 1986, \pasp, 98, 609

\bibitem[{{Hubscher}(2005)}]{2005IBVS.5643....1H}
{Hubscher}, J. 2005, Information Bulletin on Variable Stars, 5643, 1

\bibitem[{{Hubscher} {et~al.}(2005){Hubscher}, {Paschke}, \&
  {Walter}}]{2005IBVS.5657....1H}
{Hubscher}, J., {Paschke}, A., \& {Walter}, F. 2005, Information Bulletin on
  Variable Stars, 5657, 1

\bibitem[{{Kaiser}(1994)}]{1994IBVS.4119....1K}
{Kaiser}, D.~H. 1994, Information Bulletin on Variable Stars, 4119, 1

\bibitem[{{Kaiser} {et~al.}(1995){Kaiser}, {Baldwin}, {Gunn}, {Terrell},
  {Stephan}, \& {Hakes}}]{1995IBVS.4168....1K}
{Kaiser}, D.~H., {Baldwin}, M.~E., {Gunn}, J., {et~al.} 1995, Information
  Bulletin on Variable Stars, 4168, 1

\bibitem[{{Kaiser} \& {Frey}(1998)}]{1998IBVS.4544....1K}
{Kaiser}, D.~H. \& {Frey}, G. 1998, Information Bulletin on Variable Stars,
  4544, 1

\bibitem[{{Kervella} {et~al.}(2004){Kervella}, {Th{\'e}venin}, {Di Folco}, \&
  {S{\'e}gransan}}]{2004A+A...426..297K}
{Kervella}, P., {Th{\'e}venin}, F., {Di Folco}, E., \& {S{\'e}gransan}, D.
  2004, \aap, 426, 297

\bibitem[{{Kurucz}(1993)}]{1993KurCD..13.....K}
{Kurucz}, R. 1993, ATLAS9 Stellar Atmosphere Programs and 2 km/s grid.~Kurucz
  CD-ROM No.~13.~ Cambridge, Mass.: Smithsonian Astrophysical Observatory,
  1993., 13

\bibitem[{{Latham}(1992)}]{1992ASPC...32..110L}
{Latham}, D.~W. 1992, in Astronomical Society of the Pacific Conference Series,
  Vol.~32, IAU Colloq. 135: Complementary Approaches to Double and Multiple
  Star Research, ed. H.~A. {McAlister} \& W.~I. {Hartkopf}, 110

\bibitem[{{Latham} {et~al.}(1996){Latham}, {Nordstr{\"o}m}, {Andersen},
  {Torres}, {Stefanik}, {Thaller}, \& {Bester}}]{1996A+A...314..864L}
{Latham}, D.~W., {Nordstr{\"o}m}, B., {Andersen}, J., {et~al.} 1996, \aap, 314,
  864

\bibitem[{{Latham} {et~al.}(2002){Latham}, {Stefanik}, {Torres}, {Davis},
  {Mazeh}, {Carney}, {Laird}, \& {Morse}}]{2002AJ....124.1144L}
{Latham}, D.~W., {Stefanik}, R.~P., {Torres}, G., {et~al.} 2002, \aj, 124, 1144

\bibitem[{{Marsh}(1989)}]{1989PASP..101.1032M}
{Marsh}, T.~R. 1989, \pasp, 101, 1032

\bibitem[{{Nordstr{\"o}m} {et~al.}(1994){Nordstr{\"o}m}, {Latham}, {Morse},
  {Milone}, {Kurucz}, {Andersen}, \& {Stefanik}}]{1994A+A...287..338N}
{Nordstr{\"o}m}, B., {Latham}, D.~W., {Morse}, J.~A., {et~al.} 1994, \aap, 287,
  338

\bibitem[{{Olsen}(1988)}]{1988A+A...189..173O}
{Olsen}, E.~H. 1988, \aap, 189, 173

\bibitem[{{Olsen}(1994)}]{1994A+AS..106..257O}
{Olsen}, E.~H. 1994, \aaps, 106, 257

\bibitem[{{Popper} \& {Etzel}(1981)}]{1981AJ.....86..102P}
{Popper}, D.~M. \& {Etzel}, P.~B. 1981, \aj, 86, 102

\bibitem[{{Press} {et~al.}(1992){Press}, {Teukolsky}, {Vetterling}, \&
  {Flannery}}]{1992nrfa.book.....P}
{Press}, W.~H., {Teukolsky}, S.~A., {Vetterling}, W.~T., \& {Flannery}, B.~P.
  1992, {Numerical recipes in FORTRAN. The art of scientific computing}
  (Cambridge University Press)

\bibitem[{{Skrutskie} {et~al.}(2006)}]{2006AJ....131.1163S}
{Skrutskie}, M.~F. {et~al.} 2006, \aj, 131, 1163

\bibitem[{{Southworth}(2013)}]{2013A&A...557A.119S}
{Southworth}, J. 2013, \aap, 557, A119

\bibitem[{{Strassmeier} {et~al.}(2000){Strassmeier}, {Washuettl}, {Granzer},
  {Scheck}, \& {Weber}}]{2000A+AS..142..275S}
{Strassmeier}, K., {Washuettl}, A., {Granzer}, T., {Scheck}, M., \& {Weber}, M.
  2000, \aaps, 142, 275

\bibitem[{{Torres} {et~al.}(2010){Torres}, {Andersen}, \&
  {Gim{\'e}nez}}]{2010A+ARv..18...67T}
{Torres}, G., {Andersen}, J., \& {Gim{\'e}nez}, A. 2010, \aapr, 18, 67

\bibitem[{{Torres} {et~al.}(2002){Torres}, {Neuh{\"a}user}, \&
  {Guenther}}]{2002AJ....123.1701T}
{Torres}, G., {Neuh{\"a}user}, R., \& {Guenther}, E.~W. 2002, \aj, 123, 1701

\bibitem[{{Torres} {et~al.}(1997){Torres}, {Stefanik}, {Andersen}, {Nordstrom},
  {Latham}, \& {Clausen}}]{1997AJ....114.2764T}
{Torres}, G., {Stefanik}, R.~P., {Andersen}, J., {et~al.} 1997, \aj, 114, 2764

\bibitem[{{van Hamme}(1993)}]{1993AJ....106.2096V}
{van Hamme}, W. 1993, \aj, 106, 2096

\bibitem[{{VandenBerg} {et~al.}(2006){VandenBerg}, {Bergbusch}, \&
  {Dowler}}]{2006ApJS..162..375V}
{VandenBerg}, D.~A., {Bergbusch}, P.~A., \& {Dowler}, P.~D. 2006, \apjs, 162,
  375

\bibitem[{{Westera} {et~al.}(2002){Westera}, {Lejeune}, {Buser}, {Cuisinier},
  \& {Bruzual}}]{2002A+A...381..524W}
{Westera}, P., {Lejeune}, T., {Buser}, R., {Cuisinier}, F., \& {Bruzual}, G.
  2002, \aap, 381, 524

\bibitem[{{Wolf} {et~al.}(2010){Wolf}, {Claret}, {Kotkov{\'a}}, {Ku{\v
  c}{\'a}kov{\'a}}, {Koci{\'a}n}, {Br{\'a}t}, {Svoboda}, \& {{\v
  S}melcer}}]{2010A+A...509A..18W}
{Wolf}, M., {Claret}, A., {Kotkov{\'a}}, L., {et~al.} 2010, \aap, 509, A18

\bibitem[{{Wraight} {et~al.}(2011){Wraight}, {White}, {Bewsher}, \&
  {Norton}}]{2011MNRAS.416.2477W}
{Wraight}, K.~T., {White}, G.~J., {Bewsher}, D., \& {Norton}, A.~J. 2011,
  \mnras, 416, 2477

\bibitem[{{Zucker} \& {Mazeh}(1994)}]{1994ApJ...420..806Z}
{Zucker}, S. \& {Mazeh}, T. 1994, \apj, 420, 806

\end{thebibliography}
\end{document}